\documentclass[twocolumn,aps,prl,showpacs,superscriptaddress,tightenlines,a4paper]{revtex4}

\usepackage{graphicx}
\usepackage{mathrsfs}
\usepackage{caption}
\usepackage{subfig}
\usepackage{amsmath}
\usepackage{bm}
\usepackage{color}
\usepackage{pifont}

\newcommand{\p} {\prime}

\newcommand{\ome} {\omega}

\newcommand{\be}{\begin{equation}}
\newcommand{\ee}{\end{equation}}
\newcommand{\bgar}{\begin{eqnarray}}
\newcommand{\enar}{\end{eqnarray}}

\usepackage{ulem}

\begin{document}
    \title{Radiation `damping' in atomic photonic crystals}
    \date{\today}
    \author{S. A. R. Horsley}
    \email{Simon.Horsley@st-andrews.ac.uk}
    \affiliation{School of Physics and Astronomy, University of St. Andrews, North Haugh, St. Andrews, UK}
    \affiliation{Department of Physics, University of York, Heslington, York, UK}
    \affiliation{European Laboratory for Nonlinear Spectroscopy, Sesto Fiorentino, Italy}
    \author{M. Artoni}
    \affiliation{European Laboratory for Nonlinear Spectroscopy, Sesto Fiorentino, Italy}
    \affiliation{Department of Chemistry and Physics of Materials, University of Brescia, Italy}
    \author{G. C. La Rocca}
    \affiliation{Scuola Normale Superiore and CNISM, Pisa, Italy}

\begin{abstract}
        The force exerted on a material by an incident beam of light is dependent upon the material's velocity in the
        laboratory frame of reference.  This velocity dependence is known to be difficult to measure, as it is proportional to the incident optical power
        multiplied by the ratio of the material velocity to the speed of light.  Here we show that this typically tiny effect is greatly amplified in
        multilayer systems composed of resonantly absorbing atoms (e.g. optically trapped \(^{87}\text{Rb}\)), which may exhibit ultra--narrow photonic
        band gaps.  The amplification of the effect is shown to be three orders of magnitude greater than previous estimates for conventional photonic--
        band--gap materials, and significant for material velocities of a few \(\text{ms}^{-1}\).
\end{abstract}


\maketitle
    The force of radiation pressure is dependent upon the velocity of the body being pushed.  For a perfectly reflecting
    mirror this velocity dependence appears as a kinetic friction term in the equations of motion for the mirror, and 
    is due to the reduction in photon flux and frequency as observed in the material's rest frame, relative to the
    laboratory frame.  This phenomenon was predicted some time ago by Braginski and Manukin~\cite{braginski1967}, where
    it was observed that the oscillatory motion of such a mirror connected to a wall via a spring would be damped in
    proportion to the power density of the incident beam.  More recently, there was revived interest in this effect in
    reference to the precise interferometry experiments required to detect gravitational waves (e.g. the
    LIGO and Virgo projects)~\cite{matsko1996}.  However, as stated in~\cite{karrai2008}, to observe these velocity 
    dependent terms, even in the case of a perfectly reflecting metallic mirror, the most favourable parameters lead to 
    a laser power density so great that the mirror would be unlikely to remain intact.  Therefore the question must be 
    asked as to whether there are other physical systems where the fundamental velocity dependence of the force of 
    radiation pressure could possibly be observed.
    \par
    Here we examine the radiation pressure experienced by a one--dimensional multilayered atomic 
    structure~\cite{vancoevorden1996}, with incident radiation of a frequency close to an atomic transition.  This is 
    done with the help of the Maxwell stress tensor~\cite{volume8}, which enables us to arrive at an exact expression 
    for the pressure exerted by a light pulse, and in turn to asses the feasibility of measuring radiation damping with
    such an atomic structure.  For suitable choices of optical lattice period within this frequency window, an array of
    trapped \(^{87}\text{Rb}\) atoms is known to exhibit an ultra--narrow photonic--band--gap~\cite{bajcsy}, with a 
    width on the order of GHz~\cite{artoni2005, demartini}.  Due to the very high sensitivity of the optical response
    (transparency or reflection) of such a multilayer to the frequency of the incident radiation, we observe that the
    velocity dependence of the force of radiation pressure is greatly enhanced, and can take either sign.  This
    enhancement is three orders of magnitude above that recently predicted for conventional one--dimensional photonic 
    crystals~\cite{karrai2008}.  We note that our results are not specific to trapped \(^{87}\text{Rb}\), but apply 
    equally well to any system exhibiting a  photonic--band--gap on the same frequency scale.
    \vspace{2mm}

    \par
    The rate of transfer of four--momentum, \(dP^{\mu}_{\text{\tiny{MAT}}}/dt\), to a medium which is possibly 
    dispersive and absorbing, may be calculated from the electromagnetic fields in the vacuum region outside of the 
    medium,
     \begin{equation}
        \frac{dP^{\mu}_{\text{\tiny{MAT}}}}{dt}=-\int_{\partial\,\text{\tiny{MAT}}}T^{\mu j}_{\text{\tiny{FIELD}}}\,dS_{j}
        \label{fourmom},
     \end{equation}
    where the surface element \(dS_{j}\) points outward from the material surface~\cite{volume8}, and where the relevant
    components of the energy--momentum tensor are \(T^{0i}=c\epsilon_{0}(\bm{E}\times\bm{B})_{i}\) and 
    \(T^{ij}=\epsilon_{0}[\delta_{ij}(\bm{E}^2+c^{2}\bm{B}^2)/2-E_{i}E_{j}-c^{2}B_{i}B_{j}]\).  We start by considering 
    a material planar slab at \textit{rest}, upon which a beam of linearly polarized radiation of cross sectional area 
    \(A\) is normally incident. Radiation of frequency $\omega$ enters the material through the surface at \(x=x_{1}\)
    and exits through a similar surface at \(x=x_{2}\), with complex reflection and transmission amplitudes $r(\omega)$
    and $t(\omega)$.  Averaging over a time interval \(\Delta t\gg\omega^{-1}\) yields the average four--force
    experienced by the slab at rest,
    \begin{equation}
        \left\langle\frac{dP^{\mu}_{\text{\tiny{MAT}}}}{dt}\right\rangle=\frac{A\epsilon_{0} E_{0}^2}{2}\left(\begin{matrix}[1-R(\omega)-T(\omega)]\\[10pt]\left[1+R(\omega)-T(\omega)\right]\hat{\bm{x}}\end{matrix}\right),\label{rest_force}
    \end{equation}
    where \(R(\omega)=|r(\omega)|^{2}\) and \(T(\omega)=|t(\omega)|^{2}\).  Upon Lorentz transforming (\ref{rest_force}),
    one obtains the corresponding expression for the slab in \textit{motion} with velocity \(\bm{V}=V\hat{\bm{x}}\) in 
    the lab frame.  In terms of the lab (\textit{primed}) frame quantities, one has 
    \(\omega=\sqrt{(1-V/c)/(1+V/c)}\,\omega^{\prime}=\eta\,\omega^{\prime}\), \(E_{0}=\eta\,E^{\prime}_{0}\), and the 
    average four--force becomes,
    \begin{widetext}
        \bgar
        \left\langle\frac{dP^{\mu}_{\text{\tiny{MAT}}}}{dt}\right\rangle^{\prime}=
        \frac{A\epsilon_{0}(\eta E_{0}^{\prime})^2}{2\sqrt{1-(\frac{V}{c})^{2}}}
        \shoveright{\left(\begin{matrix}
            [(1-R(\eta\omega^{\prime})-T(\eta\omega^{\prime}))+\frac{V}{c}(1+R(\eta\omega^{\prime})-T(\eta\omega^{\prime}))]\\[5pt]
            [(1+R(\eta\omega^{\prime})-T(\eta\omega^{\prime}))+\frac{V}{c}(1-R(\eta\omega^{\prime})-T(\eta\omega^{\prime}))]\hat{\bm{x}}
        \end{matrix}\right)}
        \simeq
        \frac{P^{\prime}}{c}\left(\begin{matrix}\left[F^{0}_{(0)}-\frac{V}{c}F^{0}_{(1)}\right]\\[5pt]
        \left[F^{1}_{(0)}-\frac{V}{c}F^{1}_{(1)}\right]\hat{\bm{x}}\end{matrix}\right)
        \label{relativistic_force}\enar
    \end{widetext}
    where $P^{\prime}=A\epsilon_{0}c{E_{0}^{\prime}}^{2}/2$ is the incident radiation mean power.  The force exact
    expression (\ref{relativistic_force}) has a rather involved  dependence on the velocity as it stems from \(R\) and
    \(T\) that depend on the velocity through the Doppler effect.  A characteristic velocity range for cold atoms 
    experiments is such that  \(\eta\simeq 1-V/c\), so that  the reflectivity and transmissivity can be expanded to the 
    leading order in \(V/c\) around the lab frame frequency as, 
    \(R(\eta\omega^{\prime})\simeq R(\omega^{\prime})-(\omega^{\prime}V/c)\,R^{(1)}(\omega^{\prime})\), and 
    \(T(\eta\omega^{\prime})\simeq T(\omega^{\prime})-(\omega^{\prime}V/c)\,T^{(1)}(\omega^{\prime})\).  This yields the
    simpler expression on the right hand side of (\ref{relativistic_force}) where we set,
    \begin{align*}
                  F^{0}_{(0)}&=1-R(\omega^{\prime})-T(\omega^{\prime})\\
                  F^{0}_{(1)}&=1-3R(\omega^{\prime})-T(\omega^{\prime})-\omega^{\prime}\left(R^{(1)}(\omega^{\prime})+T^{(1)}(\omega^{\prime})\right)\\
                  F^{1}_{(0)}&=1+R(\omega^{\prime})-T(\omega^{\prime})\\
                  F^{1}_{(1)}&=1+3R(\omega^{\prime})-T(\omega^{\prime})+\omega^{\prime}\left(R^{(1)}(\omega^{\prime})-T^{(1)}(\omega^{\prime})\right).
    \end{align*}
    For non-dispersive materials $ F^{1}_{(0)}$ and the the first three terms in $ F^{1}_{(1)}$ are the only 
    contributions to the radiation pressure.  These are \textit{positive} and for a lossless medium of fixed 
    reflectivity $R_{0}$ reduce to the familiar results~\cite{braginski1967, matsko1996}   respectively for the pressure
    force $2 R_{0} P^{\prime}/c$ and the radiation pressure damping (friction) $ - 4 R_{0} P^{\prime}v/c^2$.  The latter 
    is typically much smaller than the former and for non-dispersive mirrors, where
    \(R(\eta\omega^{\prime})\sim R(\omega^{\prime})\) and \(T(\eta\omega^{\prime})\sim T(\omega^{\prime})\),
    velocities as large as \(V\sim10^{6}\,\text{ms}^{-1}\) are needed to observe a few per cent velocity shift,
    corresponding to a \textit{damping} contribution to the force.  For dispersive materials, on the other hand, we have 
    an extra contribution to $ F^{1}_{(1)}$ whose sign depends on the relative strength of the reflectivity and 
    transmissivity  gradients $R^{(1)}(\omega^{\prime})$ and $T^{(1)}(\omega^{\prime})$.  When \(R\) and \(T\) change 
    significantly over a frequency range \(\Delta\omega\ll\omega^{\prime}\), this contribution could be substantial, and 
    become the dominant term in $F^{1}_{(1)}$ whose sign, unlike other terms in the force expression in (3), can then 
    become negative or positive.  This amounts, respectively, to either \textit{amplification} or \textit{damping} of the
    medium relevant dynamics~\cite{karrai2008, florian-physics}.  Materials with optical reflectivity changes 
    \(\Delta R\sim1\) on the MHz range, \textit{i.e.} \(\omega^{\prime}\Delta R/\Delta\omega\sim10^{8}\), would yield an
    appreciable velocity shift in the force even at \(V\sim\,\text{ms}^{-1}\).
   \begin{figure}[h!]
          \includegraphics[width=0.4 \textwidth]{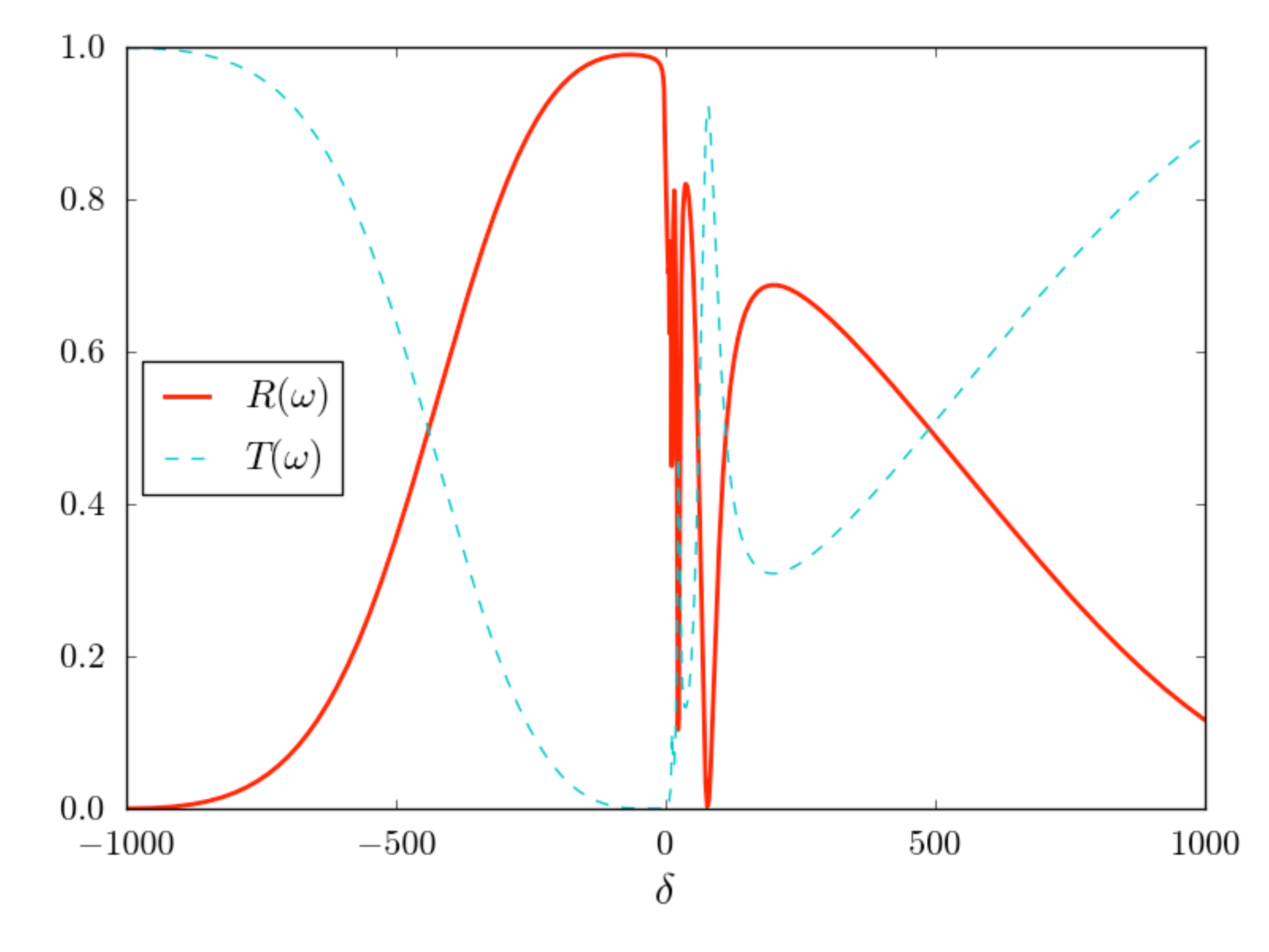}
          \caption{Reflectivity and transmissivity for a multilayered atomic structures  of \(5.3797\times10^{4}\) unit 
          cells with $a$=370.873$\times10^{-9}$ m and  $b$=19.4807$\times10^{-9}$m.  Atomic parameters are, 
          ( \(^{87}{\text{Rb}}\) $D_2$ line   \quad  $5^2 \ S_{1/2} \to 5^2 \ P_{3/2}$) \(\gamma_{e}=2 \pi \times 6 \times 10^6 \text{Hz}\),
          \(\omega_{0} = c/\lambdabar_o=2 \pi \times 384.02 \times 10^{12}\,\text{Hz}\) and an atomic density
          \(N/V=6\times10^{18}\,\text{m}^{-3}\) ( \(\mathcal{N}=1.15\times10^{-2}\)).\label{figure_R_T}}
    \end{figure}
    \par
\begin{figure}[h!]
                \includegraphics[width=0.4\textwidth]{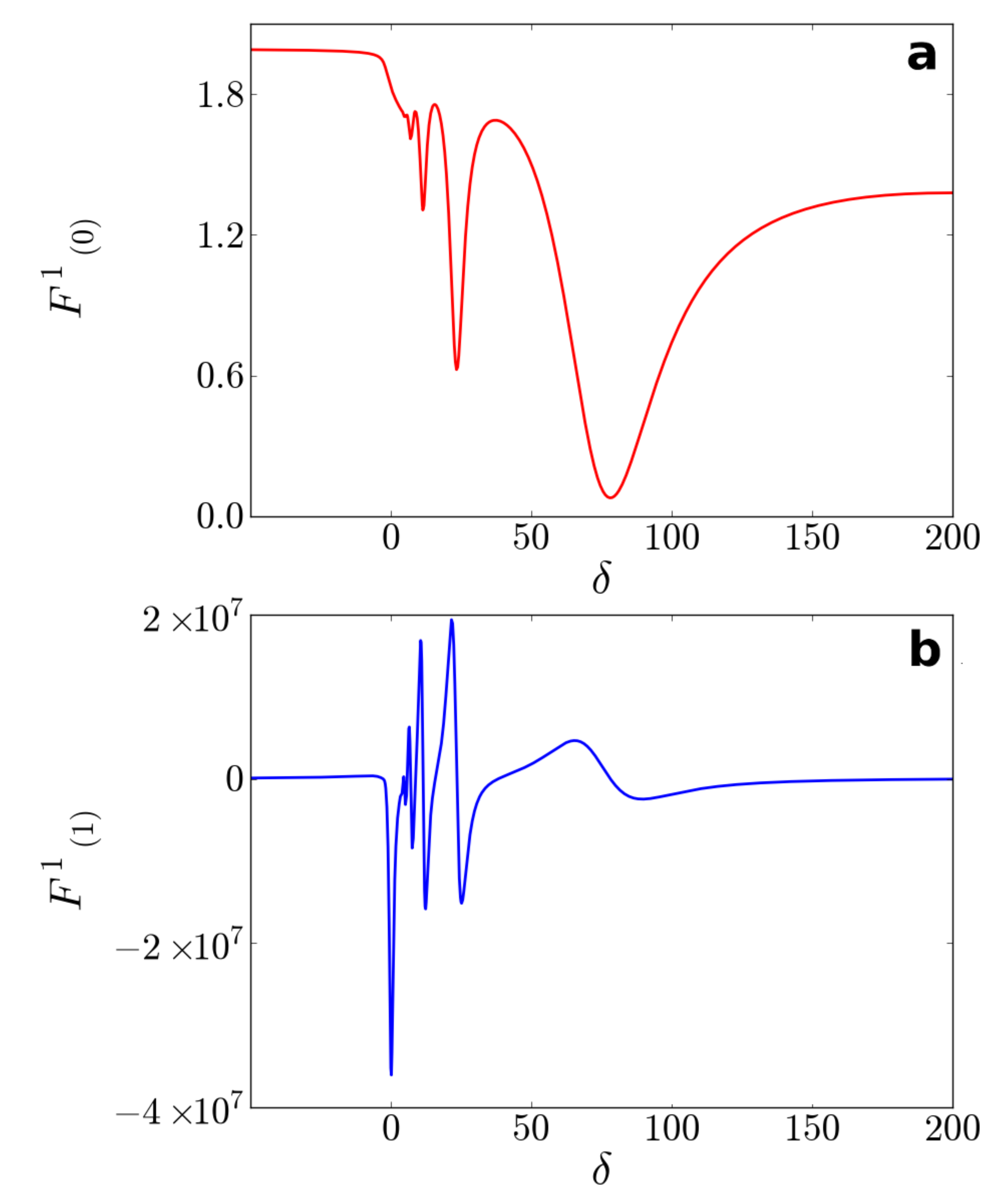}
        \caption{
                Normalised  \(x\) component of the force \(\vec{F}\) as a function of detuning around resonance where
                radiation pressure force are strongest.  The two panels show the force velocity independent
                (\(F^{1}_{(0)}\)) and velocity dependent  (\(F^{1}_{(1)}\)) parts.
         \label{figure_F}}
    \end{figure}

    Periodic structures of trapped two-level atoms separated by vacuum fit the aforementioned parameter
    range quite well~\cite{artoni2005}, and typical reflectivities and transmissivities are plotted in Fig.
    \ref{figure_R_T}  as a function of the scaled detuning $\delta=(\omega_{0}-\omega)/\gamma_{e}$ from the atomic
    resonance transition $\ome_o$, where $\gamma_e$ denotes the excited state decay rate.  These specific profiles
    correspond to a stack of alternating (complex) refractive indices~\cite{Zubairy}  $n_{a}(\omega)\simeq1$
    and $ n_{b}(\omega)\simeq \sqrt{1+3\pi\mathcal{N} /(\delta -i \, )}$ respectively with thicknesses \(a\) and \(b\),
    where $\mathcal{N}=N/(V \lambdabar_o^3)$ denotes the scaled density of atoms.  These multilayered atomic structures
    exhibit two pronounced photonic stop bands when the Bragg scattering frequency is not too far from the atomic
    transition frequency.  The first one develops from the polaritonic stop band and has one edge at the atomic 
    resonance frequency, the second one corresponds to the usual stop band of more familiar (non resonant) photonic
    crystals and develops around the Bragg frequency~\cite{artoni2005}. For the case of Fig.1 the first and second stop
    bands lie respectively on the negative and positive detuning regions.  At a given frequency $\ome^{\p}$ the 
    radiation pressure is solely determined by the multi-structure optical response and Fig. \ref{figure_F} shows both
    force components $F^1_{(0)}$ and $F^1_{(1)}$ in the red-detuned spectral region near the polaritonic stop-band in
    Fig. \ref{figure_R_T}.  In particular, Fig. \ref{figure_F}b displays the force velocity-dependent contribution,
    $F^1_{(1)}$, which largely arises from the term proportional to  $R^{(1)}(\omega^{\prime})-T^{(1)}(\omega^{\prime})$.
    This gradient difference amounts precisely to  2$\times R^{(1)}(\omega^{\prime})+A^{(1)}(\omega^{\prime})$ and,
    unlike familiar dielectric photonic crystals structures~\cite{favero2007}, absorption dispersion 
    $A^{(1)}(\omega^{\prime})$ makes an appreciable contribution to the force at the edge(s) of the polaritonic 
    stop-band where energy transfer to the sample through absorption is largest~\cite{artoni2005}.  Radiation pressure
    values can scale with \(V/c\) by as much as \(|F^{1}_{(1)}|=1\times10^{7}\) (ignoring the region very close to
    resonance) so that an appreciable ten per cent radiation pressure shift is possible with velocities of 
    \(\sim3\,\text{ms}^{-1}\), that are now within experimental reach~\cite{schmid}.  It is worth noting that the
    velocity contribution to the radiation pressure turns out to be at least three orders of magnitude larger than
    those obtained with traditional dielectric photonic crystals~\cite{karrai2008}.  In addition, depending on whether
    the difference between $R^{(1)}(\omega^{\prime})$ and $T^{(1)}(\omega^{\prime})$ is positive or negative
    alternating cooling and heating cycles become easily accessible in the appropriate spectral 
    region~\footnote{We have verified that the results do not change if the effects of a non--uniform atom density
    towards the edges of the trap are taken into account.  If \(100\) periods on each side of the trap have half the
    atom density of the centre, then a calculation of the transfer matrix shows that in the region with
    (\(|\delta|> 10\)), corrections to \(R\) and \(T\) are at most \(\sim10^{-2}\).}.
    We should note that the above analysis is approximate due to the expansion to first order in \(V/c\), yet the above
    conclusions are not fundamentally altered if all orders of \(V/c\) are retained. This is shown in the following in
    the case of a light pulse, or sequence of pulses, of finite extent rather than a (monochromatic) plane light-wave.
    This is indeed to be examined to better asses the experimental feasibility  and becomes furthermore important when
    opto-mechanical effects of radiation pressure associated with weak quantum light pulses are to be
    observed~\cite{demartini, tombesi}.  The momentum exchanged between a pulse of a specified extent and the
    multilayered atomic structure described above is here calculated by keeping all orders of \(V/c\) in the resulting
    expression.  At the position of the material surface, \(x_{1}\), the electric field of a pulse of central frequency
    \(\omega_{c}\) and half--width--half--maximum, \(\mathcal{L}\sqrt{\ln{(2)}}\), can be written as a superposition of
    its incident and reflected parts as follows (the real part of this expression entering the energy--momentum tensor),
    \begin{equation}
        \bm{E}(t)=\int_{0}^{\infty}\frac{d\omega}{\sqrt{2\pi}} \bm{\xi}(\omega)\left(e^{i\omega(x_{1}/c-t)}+
        r(\omega)e^{-i\omega(x_{1}/c+t)}\right).
        \label{fx1}
    \end{equation}
    When the pulse is linearly polarized along the \(y\)--axis and the wavelength associated with the carrier frequency
    is much shorter than the extent of the pulse (\(\omega_{c}\mathcal{L}/c\gg1\)), then a Gaussian pulse can be described
    as: \(\bm{\xi}(\omega)\simeq \sqrt{\mu_{0}\bar{N}\hbar\omega_{c}\mathcal{L}/(A\sqrt{\pi/2})}\hat{\bm{y}}e^{-(\mathcal{L}(\omega-\omega_{c})/2c)^{2}}\),
    where \(\bar{N}\) is the average number of photons within the pulse.  The transferred momentum from a pulse 
    incident on the structure entrance surface at \(x_{1}\), as observed in the material's rest frame, is calculated 
    as an integral over all time of the relevant integrated energy-momentum tensor component in (1). As an example,
    in \(T^{11}\) one encounters the integral,
    \begin{equation}
            \int_{-\infty}^{\infty}\,dt\left[\bm{E}(t)^{2}+c^{2}\bm{B}(t)^{2}\right]=\int_{0}^{\infty}d\omega\,|\bm{\xi}(\omega)|^{2}\left[1+R(\omega)\right]
    \end{equation}
    A similar contribution for the momentum transfer at the rear surface at \(x_{2}\), can be obtained starting from
    (\ref{fx1}) where we replace  \(e^{i\omega(x_{1}/c-t)}+r(\omega)\,e^{-i[\omega(x_{1}/c+t)-\phi_{r}(\omega)} \to
    t(\omega)e^{i\omega(x_{2}/c-t)}\).  The net integrated four-momentum imparted by the pulse as seen in the slab rest
    frame can be calculated and in turn transformed as a four-vector into the lab frame. If the change in the medium
    velocity is negligible over the transit time of a single pulse, the net lab (primed) frame momentum transfer per
    pulse is given by,
        \begin{equation}
            {\Delta P}^{\mu \prime}=\frac{\epsilon_{0}A\eta}{2\sqrt{1-(V/c)^{2}}}\int_{0}^{\infty}d\omega^{\prime}
            |\bm{\xi}(\eta\omega^{\prime})|^{2}\left(\begin{matrix}{\mathcal{P}^{0}}^{\prime}\\{\mathcal{P}^{1}}^{\prime}\hat{\bm{x}}\end{matrix}\right)
        \end{equation}
   where in the lab (primed) frame we have; \(\mathcal{L}=\mathcal{L}^{\prime}/\eta\);
   \(\omega_{c}=\eta\omega_{c}^{\prime}\); 
   \({\mathcal{P}^{0}}^{\prime}=1-R(\eta\omega^{\prime})-T(\eta\omega^{\prime})+(V/c)(1+R(\eta\omega^{\prime})-T(\eta\omega^{\prime}))\);
   and \({\mathcal{P}^{1}}^{\prime}=1+R(\eta\omega^{\prime})-T(\eta\omega^{\prime})+(V/c)(1-R(\eta\omega^{\prime})-T(\eta\omega^{\prime}))\).
   This recovers well know results for a transparent lossless slab~\cite{rljmo} in the limit for which $V/c \to 0$ and 
   $\eta \to 1$.  The plane wave limit in (\ref{relativistic_force}) also emerges namely when $\mathcal{L}\to\infty$,
   if $\Delta P^{\mu \prime}$ is first divided by the proper time interval for a square pulse containing the same amount
   of momentum, and of the same maximum amplitude as the incident Gaussian pulse,
   \(\Delta\tau=\sqrt{\pi/2}(\mathcal{L}/c)\).
    \begin{figure}[h!]
        \includegraphics[width=0.4 \textwidth]{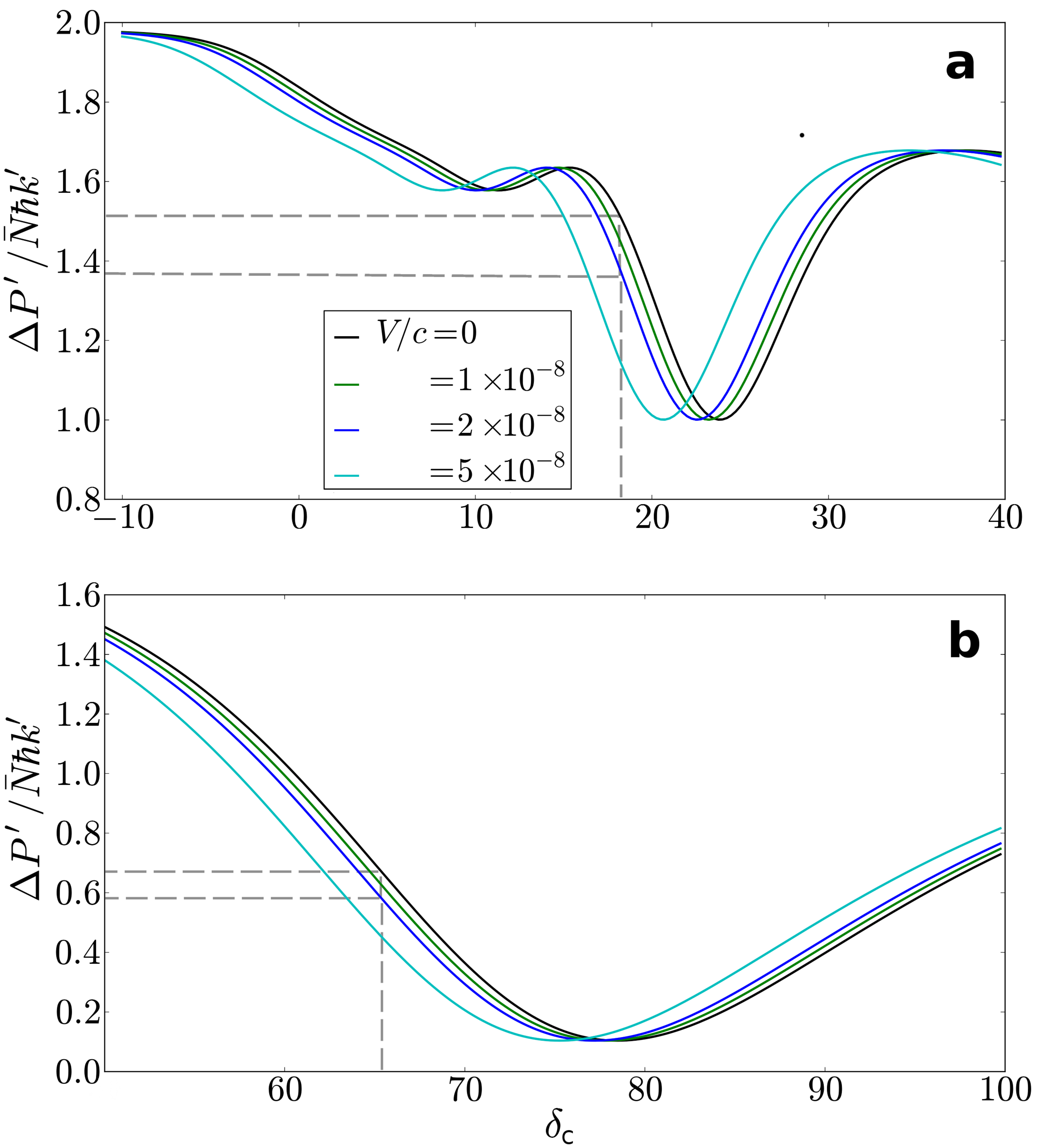}
        \caption{
        Normalised impulse, \(\Delta\vec{P}^{\prime}\cdot \hat{\bm{x}}\) acquired from a 3$m$ long Gaussian pulse impinging upon a multilayered atomic structure. The spectral region has been divided in two parts (a) and (b) for clarity.
        The detuning $\delta_c$ between the pulse carrier frequency (\(\omega_{c}\)) and atomic resonance ($\omega_{0}$) are in units of $\gamma_e$.
        All other parameters are as in fig. \ref{figure_R_T}.
        \label{figure_I}}
    \end{figure}
   The results presented in figure \ref{figure_I} illustrate the momentum transfer, 
   \({\Delta\vec{P}}^{\prime}\cdot \hat{\bm{x}}\), in units of the total momentum contained in the incident pulse in
   the lab frame, \(\bar{N}\hbar k^{\prime}_{c}\).  They substantiate the findings of the previous section: Fig.
   \ref{figure_I}a shows a difference in the radiation pressure force per pulse \(\sim10\%\) (\(\delta_c\simeq18\)) 
   relative to a stationary structure (\(V=0\)) while a similar difference of \(\sim15\%\) 
   is shown to occur away from resonance, where \(\delta_c\simeq 65\).
   \vspace{2mm}
   Radiation damping effects from multilayered atomic structures forming Bragg mirrors are best investigated in optical
   lattices created by the interference of multiple laser beams, which cool and localize atoms at the lattice 
   sites~\cite{raithel}.  For a sufficiently long trap the (Bragg) mirror may be envisaged as made of an array of disks
   spaced by half the wavelength $\lambda$ of the confining optical lattice $U(x)$ and filled with atoms in the 
   vibrational ground state of the lattice wells.  The disks thickness $d$ is essentially given by the rms position
   spread around the minima of the potential $U(x)$ while the transverse size $D\simeq \sqrt{4 N_o a/ \pi L \rho d}$ is
   determined by the in-well atomic density $\rho$ and the number of atoms $N_o$ loaded into a trap of length 
   $L$~\footnote{This is in general the case for optical traps using two counter propagating Gaussian beams.  Efficient
   radial confinement of the atoms may also be achieved by employing optical traps with Bessel beams~\cite{schmid}.}.
   For typical densities $\rho \sim 10^{12} cm^{-3}$ and filling factors $b/a \simeq 0.1$ a transverse size $D$ in the
   range $35 \div 110$ $\mu$m is obtained for  $N_o \sim 10^6 \div 10^7$ when $L\sim$ 1 cm.  This sets the incident beam
   waist $w_o$ and in turn its Rayleigh range $x_R =\pi w_o^2/\lambda_o$ that should clearly compare with the overall
   mirror length $L$.  Taking waists $w_o \sim 20 \div 90$ $\mu$m yields $x_R \sim 0.25 \div 3.2$ cm, which can safely
   be made  larger than $L$ for the cloud with more atoms. Ultracold samples of Na atoms, \textit{e.g.}, may further
   suit the case since $N_o \sim 10^8$ associated with an even shorter transition wavelength $\lambda_o$ can be
   attained~\cite{vanderStam}.

   After loading the atoms into a 1D optical lattice, whose counterpropagating beams are taken to be far (red)-detuned
   from atomic resonance~\footnote{This prevents absorption of optical lattice photons from hampering interactions
   between the light pulse and the coherent atoms during transfer of momentum.}, the lattice is further set into motion
   dragging along the atoms.  Such a motion may be achieved by changing the relative frequency detuning,
   $\Delta \omega$ of the two laser beams, which corresponds to a lattice velocity $v=\Delta \omega/ 2 k$ where $k$ is
   the average wavenumber. Through such a scheme velocities on the ms$^{-1}$ range can be reached~\cite{schmid}. Upon
   passage of a light pulse, the transferred momentum $\Delta\vec{P}^{\prime}$ induces coherent oscillations in the
   center of mass of the atomic wavepacket in the lattice wells.  This in turn induces a periodic redistribution of the
   power difference between the two counterpropagating lattice beams that can be measured~\cite{raithel}. In particular,
   every half a cycle of the atomic oscillations, the momentum of the atomic ensemble changes by
   $2\,\Delta\vec{P}^{\prime}$ and this corresponds to a coherent scattering of a number  of photons $\Delta N_{ph}$
   from one of the two counterpropagating laser beams of the optical lattice to the other which is given by
   $\hbar k \,\Delta N_{ph}=|\Delta\vec{P}^{\prime}|$. Of course, by sending a train of $N_{lp}$ light pulses equally
   spaced by the period of the atomic harmonic oscillations so to always push the atoms at the right time, 
   $\Delta N_{ph}$ would be amplified by the factor $N_{lp}$.  All-optically tailorable light pressure damping and 
   amplification effects can here be attained in the absence of a cavity, which  makes multilayered atomic structures
   not only interesting in their own right but also amenable to a new opto-mechanical regime.  The large per-photon 
   pressures that can be observed compare in fact with state of the art micro~\cite{painter-1} and
   nano~\cite{painter-2} optomechanical resonators, yet involving (atomic) masses that are various orders of magnitude
   smaller.

    \section{Acknowledgments}

   We thank G. Ferrari and L. Fallani for stimulating discussions. This
   work is supported by the CRUI-British Council Programs ``Atoms and Nanostructures''
   and ``Metamaterials'', and the IT09L244H5 Azione Integrata MIUR grant.

\bibliography{rdapc-refs}
\end{document}